\documentclass[a4paper]{article}
\usepackage{amsmath,graphicx,amssymb}
\usepackage{bm}
\pdfoutput=1

\newcommand{\logit}{\mbox{logit}}
\newcommand{\balpha}{\bm{\alpha}}
\newcommand{\btheta}{\bm{\theta}}

\begin{document}

\setlength{\textwidth}{18cm}

\title{Improved high-dimensional prediction with Random Forests by the use of co-data}
\date{}
\author{Dennis E. te Beest$^{1}$, Steven W. Mes$^{3}$, Ruud H. Brakenhoff$^{3}$, Mark A. van de Wiel$^{1,2}$}
\maketitle

\noindent
1. Department of Epidemiology and Biostatistics, VU University Medical Center, Amsterdam, The Netherlands
2. Department of Mathematics, VU University, Amsterdam, The Netherlands
3. Department of Otolaryngology-Head and Neck Surgery, VU University Medical Center, Amsterdam, The Netherlands
\vspace{8 mm}

\begin{abstract}
Prediction in high dimensional settings is difficult due to large by number of variables relative to the sample size. We demonstrate how auxiliary "co-data" can be used to improve the performance of a Random Forest in such a setting. Co-data are incorporated in the Random Forest by replacing the uniform sampling probabilities (used to draw candidate variables, the default for a Random Forest) by co-data moderated sampling probabilities. Co-data here is defined as any type information that is available on the variables of the primary data, but does not use its response labels. These moderated sampling probabilities are, inspired by empirical Bayes, learned from the data at hand. We demonstrate this co-data moderated Random Forest (CoRF) with one example. In the example we aim to predict a lymph node metastasis with gene expression data. We demonstrate how a set of external p-values, a gene signature, and the correlation between gene expression and DNA copy number can improve the predictive performance.
\end{abstract}

\vspace{4 mm}

\section*{Background}

High-dimensional prediction is inherently a difficult problem. In this paper we demonstrate how to improve the performance of the Random Forest (RF) on high-dimensional (in particular genomics) data by guiding it with `co-data'. Here, co-data is defined as any type of qualitative or quantitative information on the variables that does not use the response labels of the primary data. The primary data may, for example, be a set of gene expression profiles with corresponding binary response labels. Examples of co-data are: p-values on the same genes in a external, related study, correlations with methylation or DNA copy number data, or simply the location on the genome. Guiding a prediction model by co-data may lead to improved predictive performance and variable selection.

Several methods are able to incorporate co-data during model training. A general multi-penalty approach was suggested by \cite{tai_incorporating_2007},  a weighted lasso by \cite{bergersen_weighted_2011}, and a group-regularized ridge by \cite{van_de_wiel_better_2016}. These methods are all based on penalised regression, with a penalty parameter that is allowed to vary depending on the co-data, effectively rendering co-data based weights. The group-lasso \cite{meier_group_2008} and sparse group-lasso \cite{simon_sparse-group_2013} are also regression-based, but these methods apply a specific group-penalty that can exclude entire groups of variables. Except for the group-regularized ridge, all these methods allow for only one type of co-data. In addition, except for the weighted lasso, these methods require the co-data to be specified as groups. The weighted lasso can handle one source of continuous co-data, but requires an a priori assumption about the functional form between the penalty weighting and the co-data. For some co-data, like external p-values, this functional form is largely unknown. Hence, it may be desirable to learn it from the co-data, and to enforce monotonous weights to ensure stability and interpretability.

The Random Forest (RF) is a learner that is popular due to its robustness to various types of data inputs, its ability to seamlessly handle non-linearities, its invariance to data transformations, and its ease of use without any or much tuning \cite{breiman_random_2001}. RF are suitable and computationally efficient for genomics data, with typically the number of variables, $P$, largely exceeding the sample size, $n$ \cite{chen_random_2012,diaz-uriarte_gene_2006}. Its scale invariance makes it a good candidate to analyse RNASeq data. Due to the skewed nature of such data, their analysis is less straightforward with penalized regression techniques and results depend strongly on the data transformation applied \cite{zwiener_transforming_2014}. Our aim is to develop a co-data moderated RF (CoRF) which allows the joint use multiple types of co-data, the use of continuous co-data, and flexible modelling of the co-data weights.

The described methodology can in principle be used with any bagging classifier that uses random subspace selection, but in this paper we focus on the RF. The method is exemplified with one example. We aim to predict lymph node metastasis (LNM) for patients with head-and-neck squamous cell (HNSC) cancer using TCGA RNAseq data. We show how the use of several types of co-data, including DNA copy number, an external gene signature and mRNA microarray data from an independent patient cohort, improves the predictive performance, and validate these results on a second independent data set.

\section*{Methods}

\subsection*{Random Forest}
The aim of a supervised RF is to predict per sample $i, i=1, \ldots, n,$ an outcome $Y_{i}$ using a set of variables $X_{ij}$ where $j=1, \ldots, P$ indicates the variables. Here, we focus on binary outcome $Y_i$, although the entire methodology and software also applies to continuous and censored (e.g. survival) outcomes. A RF consists of a large number of unpruned decision trees, where each tree is grown on a bootstrap sample of the data. At each node split in each tree only a random subset of the variables are candidates, its size denoted by $mtry$, typically set at $\sqrt P$. In a standard RF, all variables have an equal probability of being candidates. Predictions are issued by majority voting across all trees. A RF is fitted to a bootstrap sample of the data implying that per tree the remaining fraction (on average 0.368) is out-of-bag (oob) and can be used to obtain an estimate of the prediction error. This leads to a computational advantage compared to methods that require cross-validation for this purpose.

\subsection*{Group specific probabilities}
We first briefly describe our method using one source of grouped co-data only. Here, the basic idea is that, when an a priori grouping of variables is available (co-data), we may sample the variables according to group-specific probabilities, and these probabilities can be estimated empirically from the data. When the number of groups is limited, only a few parameters need to be estimated (the group specific probabilities). Especially when the difference in predictive power between groups of variables is large, the predictive performance may be enhanced.

In practice, this means we first need to run a base RF (i.e. uniform sampling probabilities). From this initial fit, we obtain the number of times each variable is used across all trees. Then, the new group-specific probabilities $w_g$ are:

\begin{equation}\label{GroupRF}
w_g = (\hat{p}_g^{\normalsize{sel}} - \gamma p^0)^{+},
\end{equation}

where $\hat{p}_g^{\normalsize{sel}}$ is the proportion of selected variables from group $g$ across all trees divided by the size of group $g$ and $p^0 =1/P$ is the expected value of $\hat{p}_g^{\normalsize{sel}}$ when the group structure is uninformative. Parameter $\gamma$ can be used to tune the RF to adapt to group-sparsity by thresholding, but may also be set to one to avoid tuning. After normalizing $w_g$ such that these sum to one across variables, we obtain sampling probabilities $\tilde{w_g}$, which are then used to retrain the RF, rendering the CoRF.

\subsection*{Model based probabilities}
Next we extend the described method to allow for multiple sources of co-data, including continuous co-data. First, enumerate all node splits in all trees. Then, we define $v_{jk}$ as a binary variable indicating whether or not variable $j$ was used in the $k\textsuperscript{th}$ split, and $V_{j} = \sum_k v_{jk}$ as the total number of times that variable $j$ was used.

The main challenge in modelling multiple types of co-data, is that the various types of co-data may be collinear. We therefore need to de-tangle how well the various types of co-data explain $v_{jk}$. For that, we use a co-data model. We propose to use the logistic regression framework for this. We denote the $P \times C$ co-data design matrix by $X$, where $X_{jc}$ contains the co-data information for the $j$th variable and the $c$th co-data type, and where nominal co-data on $L$ levels is represented by $L-1$ binary co-data variables. Then, $v_{jk}$ is Bernoulli distributed with $v_{jk}\sim Bern(p_j)$, and we estimate $p_j$ using a logistic regression:

\begin{equation}\label{regreq}
\logit(p_{j}) = \alpha_0 + \sum_{c=1}^{C} X_{jc}\alpha_{c}
\end{equation}

From the co-data model, we obtain a predicted probability per variable, $\hat{p}_j$. Note that inclusion of the intercept $\alpha_0$ in (\ref{regreq}) guarantees that $\sum_{j=1}^{P} \hat{p}_j = 1$, as desired. The logistic regression establishes a marginal relationship between $v_{jk}$ and $X_{jc}$. For modelling $V_j$, first note that $v_{jk}$ contains two types of dependencies: (1) a dependency between splits $k$ for a given variable $j$, e.g. only one variable can be chosen per split; (2) dependency between variables and therefore, between their selection frequencies ($V_j$). The first dependency is addressed by using a quasi-binomial likelihood $\text{qBin}(V_j;\balpha,\tau)$ for $V_j = \sum_k v_{jk}$, which allows for an over- or underdispersion parameter $\tau$ by modelling the $\text{Var}(V_j) = \tau p_j(1-p_j)$ \cite{RBook}. We do not explicitly address the second type of dependency, which implies that the estimation is based on a pseudo-log-likelihood: 

\begin{equation}
(\hat{\balpha},\hat{\tau})=\max_{\balpha,\tau}[\sum_{j=1}^P \log(\text{qBin}(V_j; \balpha,\tau)].
\end{equation}

As a result the uncertainties of the estimates and the p-values of the co-data model do not have a classical interpretation, and cannot directly be used for inference. We are, however, primarily interested in the point estimates, which we will use to re-weigh variables:

\begin{equation}\label{weights}
w_j = (\hat{p}_j - \gamma p^0)^{+}.
\end{equation}

As earlier, $\gamma$ can be set to 1, because $p^0$ provides a natural cut-off, or $\gamma$ may be tuned to more or less sparsity. Finally, we normalise $w_j$ to obtain the sampling probabilities $\tilde{w_j}=w_j/\sum_j w_j$, which are then used to re-train the RF.

The relations we are interested in are often non-linear. E.g. for external p-values the difference between $10^{-4}$ and $10^{-2}$ may be more relevant than that between 0.25 and 0.50. We therefore extend the linear model (\ref{regreq}) to include more flexible modelling of continuous co-data with a monotonous effect, which is often natural and desirable. For that, we fit a generalised additive model with a shape contained P-spline (SCOP, \cite{pya_shape_2014}), as implemented in the \texttt{R} package \texttt{scam}. Then, equation (\ref{regreq}) becomes

\begin{equation}\label{regreq2}
\logit(p_{j}) = \alpha_0 + \sum_{c=1}^{C_1} X^{n}_{jc}\alpha_{c} + \sum_{d=1}^{C_2} f_d(X^{c}_{jd})
\end{equation}

where $X^{n}$ ($X^{c}$) denotes the sub-matrix of $X$ containing the nominal (continuous) co-data, and $f_d()$ represents a flexible function provided by the SCOP. To model $f_d$, SCOP uses $m(x)=\sum_{\ell=1}^{q} \theta_{\ell}B_\ell(x)$, where $B_\ell$ is a B-spline basis function, which is monotonously increasing when $\theta_\ell \geq \theta_{\ell-1}, \ell =1, \dots, q$ \cite{pya_shape_2014}. The monotony in $\btheta$ is enforced by defining $\btheta =\Sigma \tilde{\btheta}$, where $\tilde{\btheta} = [\tilde{\theta}_1,\exp(\tilde{\theta}_2),\ldots,,\exp(\tilde{\theta}_q)]^T$  and $\Sigma_{rs} = 0$ if $r<s$ and $\Sigma_{rs} = 1$ if $r \geqslant s$. Smoothness is enforced by penalisation of the squared differences analogous to \cite{eilers_flexible_1996}. Setting $\Sigma_{rs} = -1$ for $r \geqslant s$ renders a monotonically decreasing spline. Unrestricted splines can in principle also be used in the co-data model, but are more liable to over-fitting.

Instead of using the default of $\gamma=1$, this parameter can be tuned by a grid search. This requires calculating $w_g$  and refitting a RF for each grid-value of $\gamma$. The optimal value of $\gamma$ is then the one with the best oob performance. Possible criteria for the quality of the oob predictions are the AUC \cite{calle_auc2011}, the Brier score, or the error rate. By standard we use the AUC as this was shown to be a good indicator for the performance of a RF, also with unbalanced data \cite{calle_auc2011}. In a clinical setting most data is unbalanced and the primary interest lies in other measures such as the sensitivity and specificity (as opposed to the error rate). Note that tuning $\gamma$ with the oob predictions may result in a degree of optimism. This may be solved by embedding the procedure in a cross-validation loop. When $\gamma$ is not tuned, the oob performance of CoRF may also be slightly optimistic, because the primary data was used to estimate the weights (\ref{weights}). However, when the regression model (\ref{regreq}) is parsimonious, the overoptimism is likely small, as verified empirically in the Application section. To ensure that co-data model is parsimonious, it may be useful to perform co-data selection to removing redundant co-data sources, which also assists in interpretation. The Supplementary Material, Section 6.1, supplies an heuristic procedure to do so.

\subsection*{CoRF algorithm}
The CoRF procedure can by summarized as follows:

\begin{enumerate}
	\item Fit a base RF with uniform sampling probabilities and obtain $v_{jk}$.
	
	\item To disentangle the contributions of the various co-data sources.
	\begin{itemize}
		\item Fit co-data model (\ref{regreq}), if only linear effects are assumed.
		\item Fit co-data model (\ref{regreq2}) with shape contained P-spline(s), if flexible, monotone effects are required.
		\item Optionally: exclude redundant co-data sources and re-fit the co-data model.
	\end{itemize}
	
	\item Obtain the predicted probabilities $\hat{p}_i$ from the fitted co-data model.
	
	\item Calculate the sampling probabilities $w_j$ with threshold parameter $\gamma$. Default is to set $\gamma= 1$, optionally $\gamma$ can be tuned.
	
	\item Refit the RF for each vector of $\tilde{w_j}$.
	
	\item
	\begin{itemize}
		\item If $\gamma$ is not tuned (i.e. $\gamma$ = 1), we directly obtain the CoRF, the base RF and their oob performances.
		\item If $\gamma$ is tuned, obtain $\hat{\gamma}$ by maximising the oob performance. Tuning $\gamma$ may introduce a bias in the oob performance.
		Hence, the entire procedure is cross-validated when $\gamma$-tuning is employed.
	\end{itemize}
\end{enumerate}

%----------------------------------------------------------------------------------------------------------------------------------%----------------------------------------------------------------------------------------------------------------------------------%----------------------------------------------------------------------------------------------------------------------------------

\subsection*{Implementation}
The method as described here is implemented in a corresponding R package, called CoRF, and is available on GitHub (see Supplementary Information). It depends on the R package \texttt{randomForestSRC} for fitting the RF \cite{ishwaran2007,ishwaran2008,Ishwaran2016}. A feature of this package that is of key importance for CoRF is the option to assign sampling probabilities per variable. In addition, \texttt{randomForestSRC} applies to regression, classification and survival analysis, and by extension, so does CoRF.

For classification by the RF the recommended minimal node size is one. The node size can be tuned \cite{hastie_elements_2009}, but a RF is not very sensitive to the minimal node size. In CoRF the quality of the selected variables may influence the fit of the co-data model. Variables that are used higher up in tree are, on average, more relevant, and variables that split a node of size 2 are the least relevant. For CoRF, we believe it is better to slightly increase the minimal node size, improving the quality of the selected variables, and as a result improve the quality of the co-data model. As default in CoRF, we set the minimal node size for classification at 2.

Generally CoRF will need a larger number of trees to fit than a base RF. A base RF needs enough trees to capture the underlying signal in the data. CoRF additionally needs an indication of the relevance of each variable, which feeds back to the co-data model. Also, a co-data model that contains splines generally needs more trees than a co-data model with only linear effects. In the LNM example, described below, we set the number of trees at 15.000 to ensure convergence of both the RF/CoRF and the co-data model. A lower number of trees, e.g. 2.000, gives a similar result in terms of predictive performance, but the variability between fits increases. With CoRF we recommend to use at least 5.000 trees to ensure a reliable, good fit of the co-data model. An additional advantage of a large number of trees is that, as variability between RF fits decreases with the number of trees, when tuning, we can more reliably pick the best $\gamma$.

A RF is a computationally efficient algorithm to use with high dimensional data, primarily because at each node it selects only from $\sqrt P$ variables. CoRF inherits this efficiency and when the default $\gamma$ = 1 is used, only one RF refit is needed. Next to (re)fitting the RF, the only additional computation needed in CoRF consists of fitting the co-data model. Further tuning of $\gamma$ may improve the performance, but also requires i) refitting a RF for each value for $\gamma$, and ii) an additional cross-validation loop to assess performance, thereby increasing computational cost considerably.

The performance of RF and CoRF are evaluated by AUC \cite{calle_auc2011}, for reasons given in 2.3. Similarly to the base RF, CoRF automatically renders oob predictions. CoRF is an empirical Bayes-type classifier, which uses the relation between the co-data and the primary data to estimate sampling weights. Such double use of data could lead so some degree over overoptimism, although this will likely be limited given that the co-data model is parsimonious. In addition, when splines were used, the effective degrees-of-freedom were reduced by imposing monotony. In the example below, we verified the oob-AUC by cross-validation for the performance of CoRF on the training set.

\subsection*{Comparable methods}
To our knowledge, there is only one high-dimensional prediction method that can explicitly take \emph{multiple} sources of co-data into account: the group-regularized (logistic) ridge (\texttt{GRridge} \cite{van_de_wiel_better_2016}). CoRF provides several conceptual advantages over GRridge. First, unlike CoRF, GRridge requires discretisation of continuous co-data. Second, CoRF fits the co-data coefficients in one model, (\ref{regreq}), instead of using the co-data sources iteratively. Finally, CoRF is more computationally efficient, because it a) inherits the better computational scalability of RF with respect to $P$; and b) requires very little tuning and no iterations. Thirdly (as with a base RF), CoRF is naturally able to incorporate categorical outcomes with >2 groups. GRridge inherits the advantages of ridge regression, e.g. better interpretability of the model and the ability to include mandatory covariates. In the application section we compare the performances of these two methods for the LNM example.

 \section*{Application}

\subsection*{Predicting Lymph node metastasis with TCGA data}
To exemplify the CoRF method, we apply our method to predict a lymph node metastasis (LNM) for patients with HPV negative oral cancer using RNASeqv2 data from TCGA \cite{tcga2015}. We focus on the HPV-negatives, because these  constitute the majority (approx. 90\%) of the oral cancers, and HPV-positive tumors are known to have a different genomic etiology \cite{smeets_genome-wide_2006}. Early detection of LNM close to the site of the primary tumor is important for assigning the appropriate treatment. Diagnosis of LNM with genomic markers could potentially improve diagnosis and treatment \cite{roepman_expression_2005}.

The primary data consists of normalised TCGA RNASeqv2 profiles of head-and-neck squamous cell carcinomas (HNSCC), which were downloaded together with the matching normalised DNA copy number co-data from Broad GDAC Firehose using the R package \texttt{TCGA2STAT}. Of the 279 patients described in \cite{tcga2015}, we used the subset of 133 patients that had HPV-negative tumors in the oral cavity. Of these patients, 76 experienced a LNM and 57 did not.

To enhance the prediction of the base RF, we consider three types of co-data in this example: (1) DNA copy number; (2) p-values from the external microarray data GSE30788/GSE85446; (3) a previously identified gene signature \cite{roepman_expression_2005, roepman_multiple_2006, van_hooff_validation_2012}. These three types of co-data demonstrate the variety of co-data sources that can be included in CoRF. The DNA copy number data are measurements on the same patients. We use the cis-correlations between DNA copy number and the RNASeqv2 data. Given the nature of RNASeqv2 and DNA copy number data (discrete and ordinal, respectively), we applied Kendall's $\tau$ to calculate the correlations, giving $\tau_i, i=1, \ldots, P$. Note that the DNA data are only used during training of the predictor; these are \emph{not} required for test samples, which distinguishes this type of predictor from integrative predictors \cite{broet_prediction_2009}. The p-values of GSE30788/GSE85446 are derived from measurements of the same type of genomics features (mRNA gene expression), but measured on a different platform (microarray) than that of the primary RNAseq data and on a different set of patients. The gene signature is a published set of genes that were found to be important in a different study. Figure 1 illustrates how the various types of data are used within CoRF.

Each type of co-data has its own characteristics that needs to be taken into account in the co-data model. For the DNA copy number data, we a priori expect that a gene with a positive cis-correlation is more likely to be of importance to the tumor \cite{Masayesva2004}. We use a monotonically increasing spline $f_1$ to model the relation between $p_j$ and $X^{c}_{j1} = \tau_j$ (\ref{regreq2}).  For the p-values of GSE30788/GSE85446, we a priori expect that genes with a low p-value are more likely to important on the TCGA data, and we thus use a monotonically decreasing spline $f_2$ to model the relation between $p_j$ and $X^{c}_{j2} = \text{pval}_j$. The third type of co-data, consisting of the published gene signature is included in the co-data model (\ref{regreq}) as a binary variable: $X^{n}_{j1}=1$ when gene $j$ is part of the signature, and 0 otherwise.

Data set GSE30788/GSE85446 consists of 150 Dutch patients with a HPV-negative oral cancer tumor, and are in that respect similar to the TCGA patients. Gene expression was measured by microarray, the p-values on GSE30788/ GSE85446 were calculated with a Welch two-sample t-tests; further details on the study can be found in \cite{mes_2017}. The differences between the TCGA and the Dutch data (notably the platform and the geographical location of the patients, with immediate consequences for how patients are treated) preclude a straightforward meta-analytic data integration. Also, our focus here is on the TCGA data, which were measured on a more modern platform, but shared genomic features with the Dutch co-data may enhance the weighted predictions.

After training the base RF and the CoRF, we validate these classifiers on an independent data set (GSE84846). GSE84846 contains microarray expression data of 97 HPV-negative oral cancer patients from Italy, of whom 49 had a LNM \cite{mes_2017}. To directly apply the classifiers to the validation data, we need to account for the differences in scale between RNASeqv2 and microarray data. The RNASeqv2 data are transformed by the Anscombe transformation ($\sqrt(x+3/8)$) and both data sets are scaled to have zero mean and unit variance. We only included genes that could uniquely be matched between the two data sets (leaving 12838 genes). Since this validation does not require any re-training, the performance is directly assessed by comparing the predictions with the actual labels. As alternative to this validation, we also use the relative frequency of variables used by the base RF and CoRF on the TCGA data as sampling probabilities in training a new RF on GSE84846 data, in which case the oob-performance was used. 

We also asses the performance of the base RF and CoRF in terms of variable selection on both the training and validation data sets. For the TCGA training data set, we first select a set of genes (based on $V_{j}$), retrain on this subset, and asses the performance with a 10-fold cross-validation. For the validation data set, we first select variables on the TCGA training data with the vimp-variable-hunt as described by \cite{Ishwaran2010}. We refit with the selected set on the TCGA training data, and evaluate the performance of the refitted model on the validation data using oob-performance. To asses the stability of variable selection with a base RF and CoRF, we repeatedly (20 times) sample 84 out of 133 cases without replacement and fit a base RF and CoRF to each sampled set. Note that the sampling fraction mimics the expected fraction of independent samples in a re-sampling scheme, 0.632; we preferred subsampling over resampling here, because the latter would lead to duplicate samples in both the in- and out-of-bag samples. The overlap was assessed by calculating the average overlap between any two fits for selection sizes of $10, 20, \ldots, 100$ genes.

\subsection*{Performance on LNM example}
By examining the fit of the co-data model (Figure 2), we see that $\hat{p}_i$ is estimated higher for genes with a high cis-correlation, for genes with a low p-value on GSE30788/GSE85446, and for genes that are present in the gene signature. By prioritising these genes we observe an improvement in oob-AUC (base RF: 0.682, CoRF: 0.706, Figure 3). With 10-fold cross-validation we also see an improvement by using CoRF (cv-AUC base RF: 0.675, CoRF: 0.690). On the validation data we find a slightly larger improvement (AUC base RF: 0.652, CoRF: 0.682). Retraining on the validation data using only the sampling probabilities derived from either the base RF/CoRF fits to the TCGA data yields a similar result (oob-AUC base-RF: 0.656, CoRF: 0.695). From figure 4 we observe that CoRF also outperforms the base RF in variable selection on both the training and validation data. The stability of the gene selection, when selecting genes with the vh-vimp measure, increased on average by 17\%.  For gene selection with $V_{j}$ the stability increased by 36\%. For these data, tuning of $\gamma$ does not improve results, see supplementary material.

For comparison, we fit GRridge and the enriched RF. The global penalty parameter of the GRridge was estimated with a 10-fold cross-validation. The performance for both the GRridge and the enriched-RF on the training data was assessed with a 10-fold cross-validation. For the validation data we directly applied the resulting classifier. For GRridge, we find an cv-AUC of 0.682 on the training data, and 0.689 on the validation data. In performance this is comparable to CoRF, but note that CoRF is quicker especially if we want an estimate of the prediction error (see section 3.4).

\subsection*{Computational time}
With 5000 trees and without tuning of $\gamma$, the LNM example (n=133, P = 12838) runs in 1:18 min (single threaded on a E5-2660 cpu with 128 gb memory). By comparison, fitting a GRridge (R package \texttt{GRridge}) with a 10-fold cross-validation to estimate the global $\lambda$ takes 2:07 min for the LNM example. To estimate the predictive error by cross-validation with the GRridge these times need to be multiplied by the number of folds.

\section*{Discussion}
The LNM oral cancer cancer example demonstrates that CoRF is able to improve the base RF by using co-data. Of course, the improvement relates directly to the relevance of the co-data for the data at hand. Hence, expert knowledge on the domain and available external data is absolutely crucial with our method. Including more co-data (e.g. more information) should result in a superior performance of the selected genes. Nevertheless, we also illustrated that including non-relevant co-data usually does little harm to the performance of CoRF with respect to the base RF. 

CoRF essentially aims at reducing the haystack of genomics variables by using co-data. Of course, one could also use ad-hoc filtering methods to preselecting variables on the basis of existing information, but this introduces a level of subjectivity and sub-optimality when the threshold(s) are not chosen correctly. CoRF formalizes the weighting and thresholding process and lets the data decide on the importance of a given source of co-data. We expect CoRF to be most useful in (very) high-dimensional settings. In such settings, variables likely differ strongly in predictive ability while the size of the haystack complicates the search. In such situations our co-data approach can assist with identifying the relevant variables. For $P<n$ settings, the prediction model is trained with a (relatively small) selected set of features. This means that i) Learners not supported by co-data (e.g. base RF) are fairly well able to discriminate the important variables from the non-important ones; and ii) the small $P$ complicates good estimation of our empirical Bayes-type (sampling) weights. Hence, in such a situation, CoRF (and co-data supported methods in general) are less likely to boost predictive performance. CoRF is weakly adaptive in that it learns the sampling weights from both the primary and the co-data, in contrast to other adaptive methods like the enriched RF \cite{amaratunga_enriched_2008} or the adaptive lasso \cite{ZouAdaptiveLasso} where weights are inferred only from the primary data. In high-dimensional applications such strong adaptation is more likely to lead over-fitting, unlike the co-data moderated adaptation.

CoRF inherits its computational efficiency from the RF. When the tuning-free version is used ($\gamma=1$), we empirically found that the oob performance suffices and cross-validation is not required. This makes the methodology very suitable for applications with extremely large $P$. Tuning of $\gamma$ may slightly improve the predictive performance, but at a substantial computational cost, given the required grid search for $\gamma$ and the additional CV loop. The CoRF methodology may be combined with any bagging classifier that uses random subspace selection, such as a random glm \cite{songrandom2013} or a random lasso \cite{wang_random_2011}. If variable selection is more stringent for a particular method (i.e. less noisy), then identification of the relationships of the co-data model may be easier. On the other hand, if most of the variables are not used, then we are unable to obtain an reliable assessment of the quality of those variables which may complicate fitting the co-data model. One possible improvement for CoRF could be to use the depth at which variables are used by the RF, for example through the average or minimal depth \cite{Ishwaran2010}. Variables that are used higher up in a tree are, on average, more relevant, and it could be beneficial to give these variables, for example, a bigger weight in the co-data model. Another way of accomplishing this could be by replacing $v_{ij}$ by a measure that counts how often each variable is used in classifying the oob samples, analogous to the IPS measure \cite{Epifanio, Epifanio2}, which naturally gives more weight to variables that often high up in a tree. We intend to investigate these matters in the future.

\section*{Software}
The R package \textit{CoRF} is available freely from GitHub: https://github.com/DennisBeest/CoRF.

\section*{Competing interests}
The authors declare that they have no competing interests.

\section*{Authors' contributions}
DB and MW developed the method, DB developed the R code, applied it to the example and drafted the manuscript. MW revised the manuscript. SM and RB provided input for the LNM example.  All authors read and approved the final manuscript.

\section*{Acknowledgements}
This study was supported by the OraMod project, which received funding from the European Community under the Seventh Framework Programme, grant no. 611425. Data sets GSE30788, GSE85446, GSE84846 were pre-processed in the OraMod project. The results of LNM example are partly based upon data generated by the TCGA Research Network.

\bibliography{CoRFrefs}
\bibliographystyle{unsrt}

\newpage

 \begin{figure}[!ht]
	\includegraphics[width=5in]{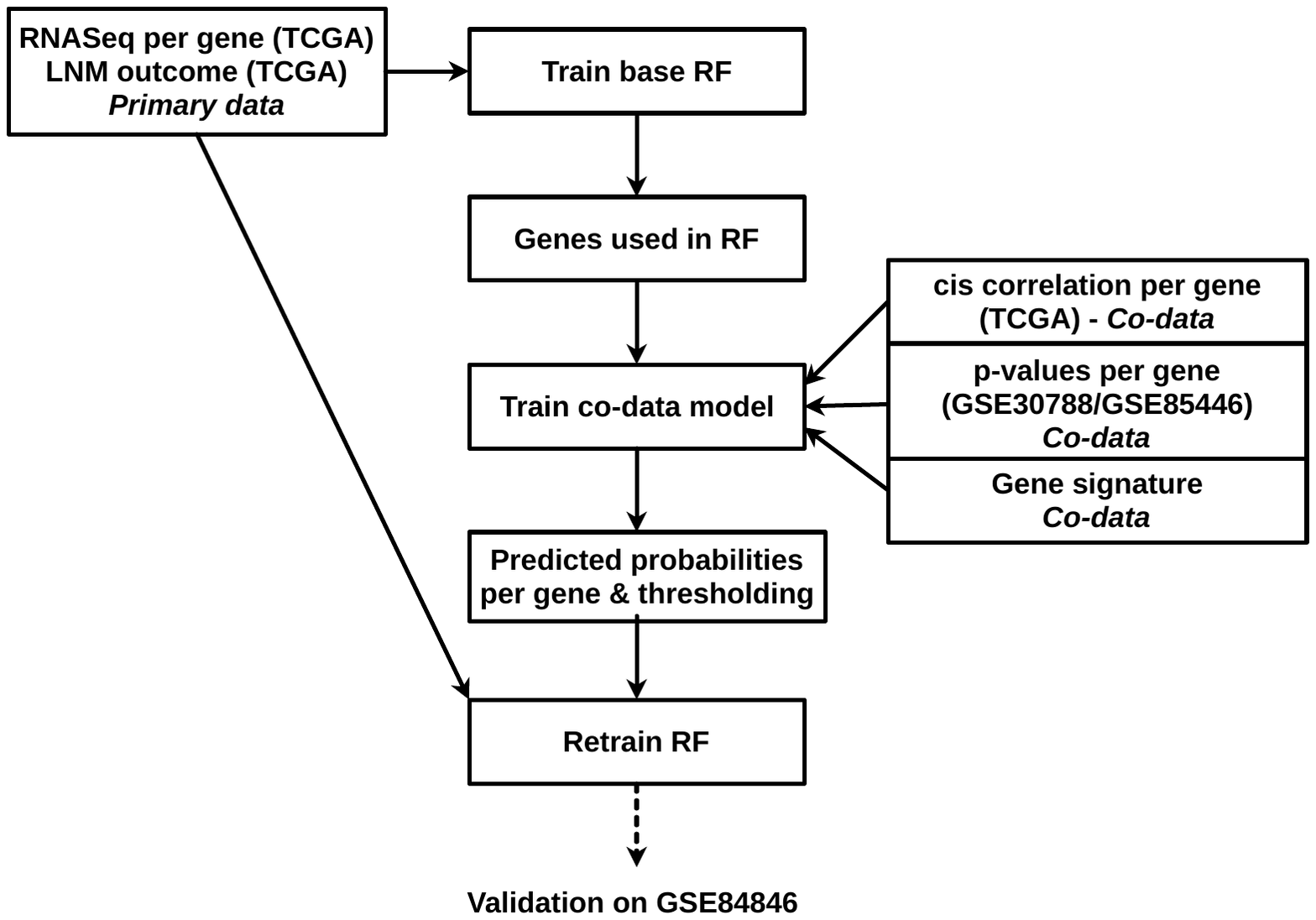} % uncomment to see figure
	\caption{Illustration of the sources of data used in CoRF for the LNM example. First a base RF is fitted on the training data. Its output, $v_{ij}$, together with the co-data, is used to train the co-data model. From the co-data model, we obtain a probability per gene used for refitting on the training data. In an extra step we validate the results on GSE84846.}
 	\label{figure_1}
 \end{figure}
\newpage
\begin{figure}[!ht]
	\includegraphics[width=4.5in]{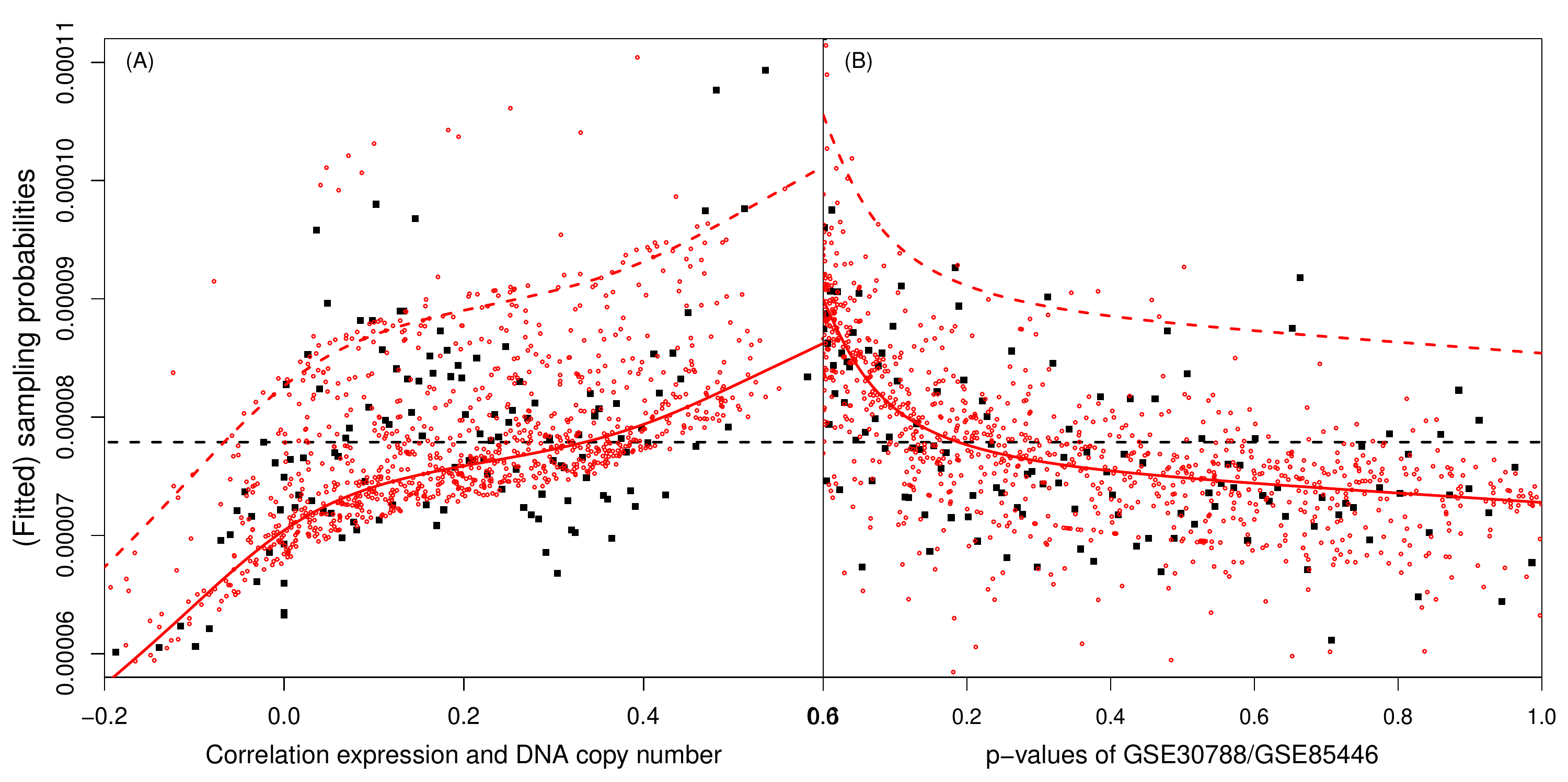} % uncomment to see figure
	\caption{Fit of the co-data model for the LNM example.  Each square represents 100 genes grouped by either (A) DNA copy number-expression correlation or (B) p-value. The red lines represent the marginal fit across the correlations or p-values. The top red lines represent the fit for genes present in the gene profile. The cloud of red dots represent the fitted values for 1000 randomly selected genes.}
	\label{figure_2}
\end{figure}
\newpage

\begin{figure}[!ht]
	\includegraphics[width=4.5in]{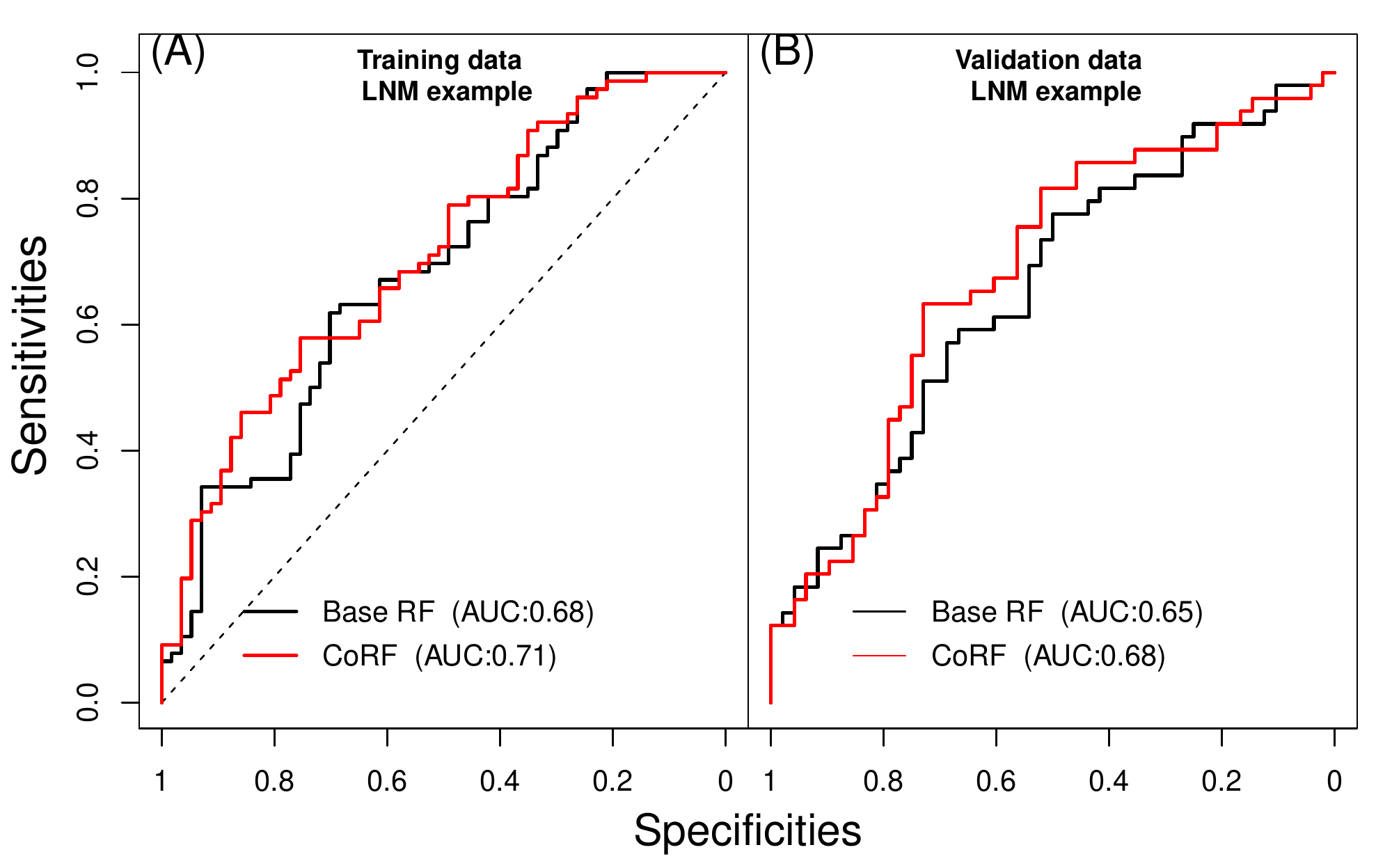} % uncomment to see figure
	\caption{The ROC curve based on oob predictions for the base RF and CoRF. The ROC curve based on oob predictions for the base RF and CoRF; (A) the TCGA training data, (B) validation data set (GSE84846).}
	\label{figure_3}
\end{figure}
\newpage
\begin{figure}[!ht]
	\includegraphics[width=4.5in]{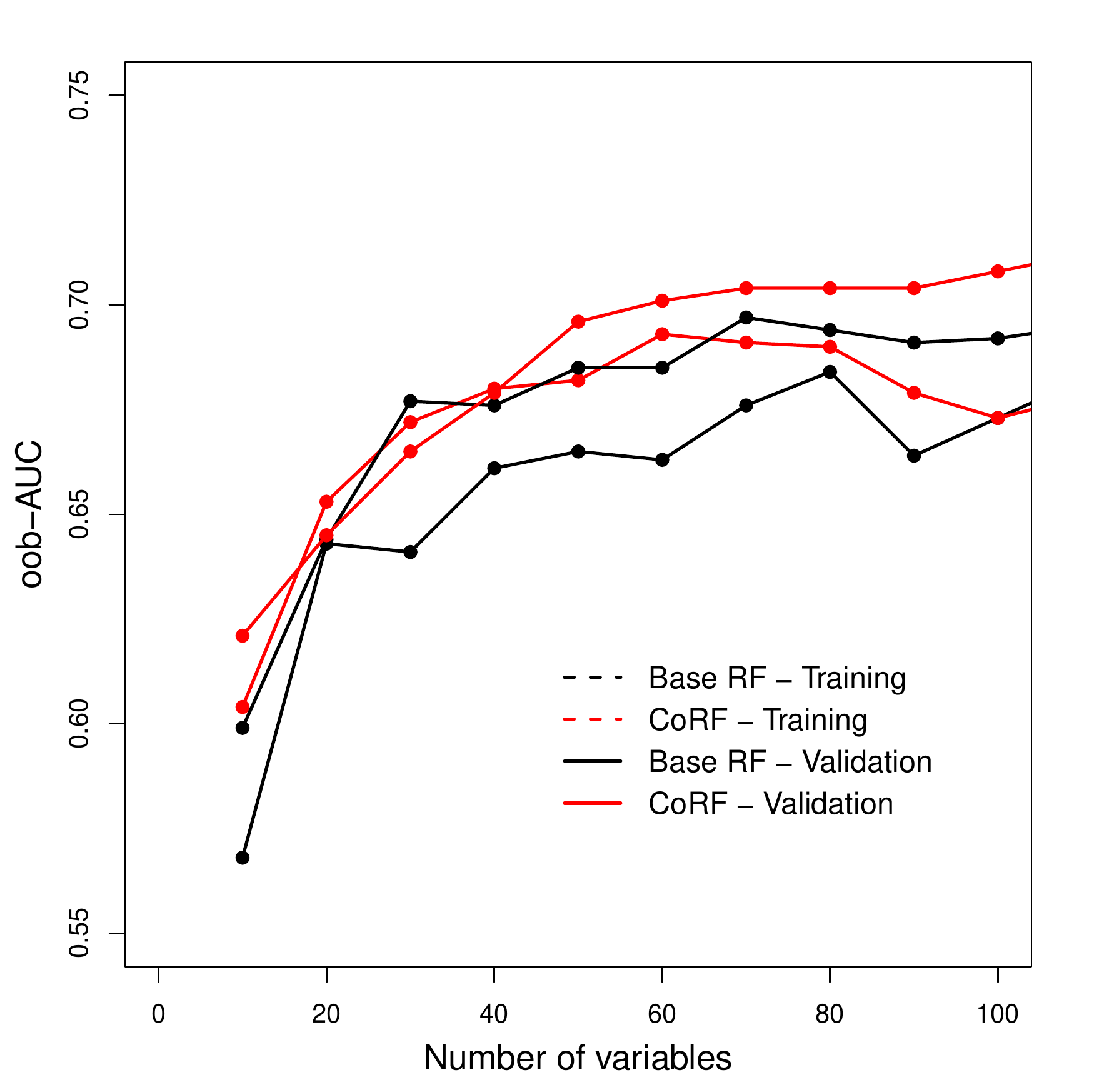} % uncomment to see figure
	\caption{The performance of RF/CoRF for given numbers of variables selected with vh-vimp for the LNM example. For the (TCGA) training data the performance was assessed by a 10-fold cross-validation. For the validation data set (GSE84846) the prediction models where directly applied.}
	\label{figure_4}
\end{figure}

\section*{Additional Files}
Supplementary material.

\end{document}